\documentclass[useAMS,usenatbib]{mn2e}
\usepackage{graphicx}
\usepackage{amssymb}
\usepackage{amsmath}
\usepackage{times}
\usepackage{epsfig}

\title[Monthly Notices] {Misconceptions About General Relativity in
  Theoretical Black Hole Astrophysics}
  \author[D. Garofalo et al.]
  {D.~Garofalo$^1$\thanks{david.a.garofalo@jpl.nasa.gov} \&
  D.L.~Meier$^1$\\
  $^1$Jet Propulsion Laboratory, California Institute of Technology,
      Pasadena, CA 91109, USA}
\pagerange{\pageref{firstpage}--\pageref{lastpage}} \pubyear{2010}

\def\LaTeX{L\kern-.36em\raise.3ex\hbox{a}\kern-.15em
    T\kern-.1667em\lower.7ex\hbox{E}\kern-.125emX}

\begin{document}

\label{firstpage}

\maketitle

\begin{abstract}
The fundamental role played by black holes in our study of
microquasars, gamma ray bursts, and the outflows from active galactic
nuclei requires an appreciation for, and at times some in-depth
analysis of, curved spacetime.  We highlight misconceptions
surrounding the notion of coordinate transformation in general
relativity as applied to metrics for rotating black holes that are
beginning to increasingly appear in the literature.  We emphasize that
there is no coordinate transformation that can turn the metric of a
rotating spacetime into that for a Schwarzschild spacetime, or more
generally, that no coordinate transformation exists that can
diagonalize the metric for a rotating spacetime.  We caution against
the notion of ``local'' coordinate transformation, which is often
incorrectly associated with a global analysis of the spacetime.
\end{abstract}

\begin{keywords}
\ relativity - rotating black holes\ 
\end{keywords}

\section{Introduction}

Not surprisingly, the overwhelming observational evidence which has
propelled black holes center stage in astrophysics over the past few
decades, has had the consequence of making theoretical aspects of
general relativity and the curved spacetime around black holes,
indispensable tools.  As a result, theoretical research in such areas
can require a mastery of apparently disparate topics ranging from
magnetohydrodynamics of turbulent flows to differential geometry, a
daunting task that is not easily accomplished well.  Given this state
of affairs, it is perhaps not surprising that misconceptions arise.
In particular, the spacetime of a rotating, isolated, black hole, in
which commonly used coordinate systems suffer singularities at the
horizon and complications due to off-diagonal terms, have been studied
via the introduction of new coordinates that simplify calculations and
avoid singularities.  Because the metric tensor is usually cast in its
one-form expression in terms of basis one-forms or differentials,
differentials of coordinate transformations are needed to recast the
metric in the new coordinate system.  Often times, this differential
approach in recasting the metric tensor is undertaken directly without
care to ensure the validity of the actual coordinate transformation.
This misconception has appeared in papers and books in the last
several years, causing confusion and erroneous results.  In this short
paper, we describe this coordinate transformation misconception and
ways to avoid it.

\section{Coordinate transformations for rotating spacetimes:the Boyer-Lindquist case}

In commonly used Boyer-Lindquist coordinates, the metric for a
rotating black hole in the coordinate basis assumes its standard form
(e.g. Poisson 2004),
\begin{eqnarray}
  dS^{2}&=&-\left(1-\frac{2Mr}{\rho^{2}}\right)
  dt^{2}-\frac{4Mar\sin^{2}\theta}{\rho^{2}}dt\,d\phi \nonumber\\
  &+&\frac{\Sigma}{\rho^{2}}\sin^{2}\theta\, 
  d\phi^{2}+\frac{\rho^{2}}{\Delta}dr^{2}+\rho^{2}d\theta^{2},
\label{Kerr}
\end{eqnarray}
where 
\begin{equation}
\rho^{2} = r^{2} + a^{2}\cos^{2}\theta,
\end{equation}
\begin{equation}
\Delta = r^{2}-2Mr+a^{2},
\end{equation}
and
\begin{equation}
\Sigma = (r^{2}+a^{2})^{2}-a^{2}\Delta\sin^{2}\theta,
\end{equation}
which can be recast in the following form, using the same coordinates.
\begin{eqnarray}
  dS^{2}&=&-\left(\frac{\rho^{2}\Delta}{\Sigma}\right)
  dt^{2}+\frac{\Sigma\sin^{2}\theta}{\rho^{2}}(d\phi - \omega dt)^{2}\nonumber\\
  &+&\frac{\rho^{2}}{\Delta}dr^{2}+\rho^{2}d\theta^{2},
\label{diagonal}
\end{eqnarray}
where
\begin{equation}
\omega = \frac{2Mar}{\Sigma}.
\label{omega}
\end{equation}
From the above form, it is easy to see that if one allows the
differential of a new coordinate $\phi'$ to have the form
\begin{equation}
d\phi' = d\phi - \omega dt,
\label{dphi}
\end{equation}
the second term in the above metric tensor (colloquially referred to
as the line element) fails to generate off-diagonal metric terms and
one is tempted to claim that a global coordinate transformation from
$\phi$ to $\phi'$ exists which accomplishes this (Krolik 1998).  This
is a misconception that arises from the differential approach.  The
coordinates ($t$, $r$, $\theta$, $\phi'$) are valid {\em local}
coordinates (in the same manner as spacetime is locally Minkowskian),
but they are not valid {\em global} coordinates.  Let us illustrate
this by going back to the actual coordinate transformation that is
implicit in the one-form of equation \ref{dphi} which is
\begin{equation}
\phi' = \phi -\omega t.
\end{equation}
From here we calculate the differential or one-form of $\phi'$ with
respect to the coordinates $r$, $\theta$, $\phi$ and $t$, via
\begin{equation}
d\phi'=\frac{\partial \phi'}{\partial r}dr + \frac{\partial
\phi'}{\partial \theta}d\theta + \frac{\partial \phi'}{\partial \phi}d\phi
+ \frac{\partial \phi'}{\partial t}dt.
\end{equation}
Therefore,
\begin{equation}
d\phi'= -\frac{\partial \omega}{\partial r}t dr - \frac{\partial
\omega}{\partial \theta}t d\theta +d\phi -\frac{\partial
\omega}{\partial \phi}td\phi -\omega dt.
\end{equation}
If $\omega$ is not a function of $\phi$, we have
\begin{equation}
d\phi'= -\frac{\partial \omega}{\partial r}t dr - \frac{\partial
\omega}{\partial \theta}t d\theta +d\phi -\omega dt,
\end{equation}
and only when $\omega$ is neither a function of $r$ nor $\theta$ does
the differential reduce to
\begin{equation}
d\phi'= d\phi -\omega dt.
\end{equation}
This means that if $\omega$ has a value at some $r$, $\theta$, such
that the metric is diagonal there, it will be diagonal only at that
point.  Therefore, no global analysis of the spacetime can be carried
out with a diagonal metric (as in Lyutikov 2009) because such an
analysis is only valid at one value of $r$ and $\theta$.  It is true
that along the worldline of an observer for which
$\frac{d\phi}{dt}=\omega$, no $d\phi'$ term exists.  However, this is
a local statement that refers to the values of $r$ and $\theta$
followed by this particular observer in spacetime.  The problem with
Krolik (1998) is in the statement that ``the two metrics agree'' when
referring to the coordinate transformation between the Kerr metric in
Boyer-Lindquist coordinates and the metric in ``locally non rotating
frame coordinates''.  It would be better to emphasize the
``agreement'' between the metric in Boyer-Lindquist coordinates and an
infinite number of {\em different} ``locally non rotating frame
coordinate'' metrics, one for each of the different values of the
coordinates $r$ and $\theta$.

Alternatively, one could take a more rigorous and possibly more direct
approach by considering the metric as a (0,2) tensor resulting from
the tensor product of one-forms as in
\begin{equation}
g = \sum_{i,j}^{}g_{ij}(dx^{i}\otimes dx^{j}).
\label{g}
\end{equation}
Accordingly, $d\phi'$ is the one-form basis in terms of unprimed
coordinates that enters the metric via equation (\ref{g}), which would
have the following expression
\begin{eqnarray}
  g=-\left(\frac{\rho^{2}\Delta}{\Sigma}\right)
  (dt \otimes dt)+\frac{\Sigma\sin^{2}\theta}{\rho^{2}}(d\phi'\otimes d\phi')\nonumber\\
  +\frac{\rho^{2}}{\Delta}(dr\otimes dr)+\rho^{2}(d\theta \otimes d\theta).
\label{g_tensor}
\end{eqnarray}

However, in order to cast $g$ in the form above, there must be a
one-form basis $d\phi'$ that is independent of $r$ and $\theta$ to
take expression (\ref{diagonal}) into expression (\ref{g_tensor});
but, by virtue of the fact that $\omega=\omega(r,\theta)$ via
expression (\ref{omega}), no such form exists.

Despite the absence of a diagonal metric, it is still possible to
avoid the complications due to off diagonal metric terms by employing
the methods of local frames or the tetrad formalism.  This simplifies
calculations by projecting tensors onto a local orthonormal basis of
four linearly independent vector fields, the frames of locally
non-rotating obsevers (Bardeen et al 1973; Chandrasekhar 1992), as
done, for example, in Shafee et al (2008), via the use of the stress
tensor component in the orthonormal basis of the comoving fluid.  Even
within the tetrad formalism, however, the Kerr metric in
Boyer-Lindquist coordinates remains of the form (\ref{Kerr}) or
(\ref{diagonal}) and is in general non-diagonal in any coordinates.
In the next section we briefly illustrate this by showing how the
introduction of one set of coordinates (which avoids the coordinate
singularity at the horizon) inevitably produces off-diagonal terms in
the spatial 3-metric.

\section{Coordinate transformations for rotating spacetimes:the Kerr-Schild case} 

In this section we perform a coordinate transformation that avoids the
coordinate singularity at the black hole horizon, emphasizing the role
of its global nature in the failure of this transformation to produce
a diagonal metric. The coordinate singularity appears in the term
\begin{equation}
  g_{rr}=\frac{\rho^{2}}{\Delta},
\label{grr}
\end{equation}
whose denominator goes to zero at the horizon.  To perform a
singularity-removing coordinate transformation requires transformation
of both time and azimuthal angle in the form
\begin{equation}
dt' = dt + \frac{2Mr}{\Delta}dr
\end{equation}
and
\begin{equation}
d\phi' = d\phi + \frac{a}{\Delta}dr.
\end{equation}
Given the fact that each second term is integrable in $r$, this
constitutes a global coordinate transformation.  Integrating $t'$, 
and $\phi'$ gives the actual coordinate transformations as
\begin{eqnarray}
t' = t +
\frac{M}{\sqrt{M^{2}-a^{2}}}[(M+\sqrt{M^{2}-a^{2}})\ln|\frac{r}{M+\sqrt{M^{2}-a^{2}}}-1|\nonumber\\
-(M-\sqrt{M^{2}-a^{2}})\ln|\frac{r}{M-\sqrt{M^{2}-a^{2}}}-1|]
\label{t'}
\end{eqnarray}
and
\begin{eqnarray}
\phi' = \phi +
\frac{a}{2\sqrt{M^{2}-a^{2}}}\ln|\frac{r-(M+\sqrt{M^{2}-a^{2}})}{r-(M-\sqrt{M^{2}-a^{2}})}|.
\label{fi}
\end{eqnarray}
In coordinates ($t'$, $r$, $\theta$, $\phi'$) the metric now becomes
\begin{eqnarray}
dS^{2}&=&-\left(1-\frac{2Mr}{\rho^{2}}\right)
dt'^{2}-2\frac{\omega\Sigma\sin^{2}\theta}{\rho^{2}}dt'\,d\phi'\nonumber\\
&+&\frac{4Mr}{\rho^{2}}dt'dr-2a\sin^{2}\theta(1+\frac{2Mr}{\rho^{2}})drd\phi'+\frac{\Sigma}{\rho^{2}}\sin^{2}\theta\,
d\phi'^{2}\nonumber\\&+&(1+\frac{2Mr}{\rho^{2}})dr^{2}+\rho^{2}d\theta^{2}.
\end{eqnarray}
With the above coordinate transformations (\ref{t'})/(\ref{fi}), the
spatial part of the metric no longer suffers a coordinate singularity
at the black hole horizon where $\Delta=0$.  However, their is a price
to pay, as the spatial 3-metric is no longer diagonal due to a new
metric term $g_{r\phi'}$.  Any attempt to produce horizon-penetrating
coordinates that avoid the coordinate singularity at the horizon,
while also leaving the spatial 3-metric diagonal, requires that the
integrand in $t'$ be non-integrable in $r$. Also, in general,
producing a diagonal metric requires the integrand in $\phi'$ be
non-integrable in $r$ as well.

\section{Discussion and Conclusions}

The property of rotating spacetimes discussed above can strongly
affect how codes that perform numerical simulations of accretion onto
black holes are constructed.  For example, Koide (2003) devised a
clever method of performing numerical MHD simulations in a stationary
Kerr metric using standard methods of curvilinear coordinates, if the
spatial portion of the 4-metric were diagonal.  This produced
excellent results outside the horizon, but, as explained above,
possessed a coordinate singularity at the horizon.  Hence, matter
could not flow naturally through the horizon in the simulation.  In
such cases, simulations could last for only a few tens of black hole
light crossing times.  On the other hand, McKinney (2006), for
example, employed the correct non-diagonal Kerr-Schild coordinates in
his simulation code, which allowed one to follow the flow of plasma
well into the horizon without numerical problems.  Such simulations
can be run out to at least $\sim 10^{4}$ black hole light crossing
times.  However, obtaining this capability in one's simulation code
requires using more involved techniques to handle the inevitable
off-diagonal 3-metric terms.

In this short letter, we highlight the increased appearence in the
astrophysical literature of the misuse of a metric of the form
(\ref{g_tensor}) when working with asymptotically flat, stationary and
axisymmetric spacetimes to address the global spacetime nature.  We
suggest that avoiding such problems can be accomplished by either
starting with the full coordinate transformation before determining
its differential form, or, by simply ensuring that the partial
derivatives in the differential are taken with respect to the full set
of coordinates.

\section{acknowledgments}

D.G. thanks the anonymous referee for minor but useful comments,
Andrew King for specific suggestions that have improved the impact of
this work, William Goldman and Sean Carroll for reading an early draft
and Ted Jacobson for past discussion on this topic.  The research
described in this paper was carried out at the Jet Propulsion
Laboratory, California Institute of Technology.

\section*{References}

\noindent Bardeen, J.M., Press, W.H., Teukolsky, S. A., 1972, ApJ, 178, 347

\noindent Chandrasekhar, S., 1992, {\em The Mathematical Theory of
Black Holes}, Clarendon Press, Oxford

\noindent Koide, S., 2003, PhysRevD, 67,104010 

\noindent Krolik, J.H., {\em Active Galactic Nuclei: From the Central
Black Hole to the Galactic Environment}, Princeton University Press,
1998.

\noindent Lyutikov, M., 2009, MNRAS, 396, 1545

\noindent McKinney, J.C., 2006, MNRAS, 368, 1561 

\noindent Poisson, E., 2004, {\em A Relativist's Toolkit, Cambridge
University Press}, UK

\noindent Shafee R., McKinney J.C., Narayan R., Tchekhovskoy A.,
Gammie C.F. \& McClintock E., 2008, ApJ, 687, L25

%%\begin{thebibliography}

%%\end{thebibliography}

\end{document}